# UMA SEQUÊNCIA DIDÁTICA CRIATIVA PARA ENSINAR AS LEIS DE NEWTON

## A CREATIVE TEACHIN SEQUENCE TO TEACH NEWTON LAWS

Ayler F. Horta Oliveira[1], Lucia Helena Horta Oliveira[2]

[1] Universidade Federal do Espirito Santo

[2] Secretaria do Estado do Espirito Santo

**Resumo:** A realidade tecnológica e digital tem se expandido, atingindo fortemente os espaços escolares. Esta pesquisa tem por objetivo desenvolver uma sequência didática, utilizando, como estratégias uma história em quadrinhos, figuras, gifs e vídeos, programadas para serem aplicadas às Leis de Newton. A pesquisa visou a analisar de que forma essa ferramenta pedagógica pode contribuir para a aprendizagem do aluno. Trata-se de uma pesquisa qualitativa em educação, cujos dados foram colhidos no transcurso da ação pedagógica, por meio de questionários, analisados comparativamente. As estratégias tiveram boa aceitação entre os adolescentes, razão pela qual sua aplicação favoreceu avanços na aprendizagem de conceitos e na aplicação de conhecimentos da Física clássica nos eventos do cotidiano.

Palavras-chave: Gifs no ensino de Física. Realidade aumentada. Quadrinhos. Mecânica clássica.

**Abstract:** The technological and digital reality has broadly expanded, reaching the school. Therefore, this research aims to develop a didactic sequence, using as strategy a comic book, Figures, Gifs and videos, programs for activities applied to comics using Newton's Laws. The research aimed to analyze how this pedagogical tool contributed to student learning. It is a qualitative research in education, whose data were collected in the course of the pedagogical action, through questionnaires, analyzed comparatively. The strategies were well accepted among adolescents, which is why their application favored advances in the learning of concepts and in the application of knowledge of classical physics in everyday events

Keywords: Gifs in physics education. Augmented Reality. Comics. Classic mechanics.

## 1 INTRODUÇÃO

A realidade tecnológica e digital tem atingindo diversos contextos sociais, sobretudo a escola. As últimas gerações de alunos, potencialmente imersas na era digital, impõem às escolas novos desafios, tais como a reformulação de métodos de ensino e preparo de professores. O ensino tradicional de Física possui um alto grau de complexidade, dificultando a relação entre o conteúdo ministrado e o contexto social do aluno. Decorre daí o desinteresse e a dificuldade de assimilação da disciplina. O ensino de Física e as dificuldades dos educandos na compreensão de conceitos vêm sendo objeto de estudo ao longo dos anos. Carvalho e Lemos (2010), Camargo *et al*. (2010) apontam para um mesmo fator potencializador dessa dificuldade – o modo como a Física é apresentada e ensinada aos alunos.

Diversas ferramentas podem auxiliar no processo de ensino-aprendizagem do educando, como as tecnologias multimídia, por exemplo, cuja utilização cresce no ambiente escolar (DONZELLI; TOMAZELLO, 2016). A inclusão de recursos digitais e tecnológicos é garantida, também, pelos Parâmetros Curriculares Nacionais (PCN's) dando amparo ao professor para explorar tais recursos de maneira livre. O enfoque desta pesquisa aponta para o uso da realidade aumentada, por meio de gifs e vídeos, no ensino de Física. Assim, este artigo buscou desenvolver uma sequência didática, utilizando-se de quadrinhos, gifs e 3D com programação de vídeos, e realidade aumentada, de modo a abordar as Leis de Newton.

## 2 REFERENCIAL TEÓRICO

A mediação, para Vygotsky, é central em sua psicologia (VYGOTSKY, 1930). De modo amplo, pode-se entender a mediação como um processo de intervenção, em que se insere um elemento, em uma relação, a qual deixa de ser direta, passando a ser mediada pelo elemento. Instrumentos, signos, ferramentas psicológicas ou seres humanos podem favorecer a mediação do aluno com o mundo. Para o psicólogo russo, nossa relação com o mundo não é direta, mas mediada e os instrumentos contribuem para tal mediação, funcionando como estimulantes de mudanças externas, pois estes tornam possível a intervenção na natureza. Rego (2001) afirma que os seres humanos têm a capacidade de elaboração e produção de seus próprios instrumentos, que resultarão na realização de tarefas específicas sendo que os indivíduos podem transmitir a função destes para outros membros de suas comunidades, recriando, e aprimorando, tais instrumentos.

Quanto à mediação pedagógica, o processo de intervenção busca ajudar o aluno a ser capaz de construir seus conhecimentos, formar novos conceitos a respeito do mundo e sobre ele intervir. A mediação seria um processo de construção de significados, cujo objetivo é ampliar as possibilidades de diálogo, de modo a desenvolver criticamente processos e conteúdos explorados nos espaços educacionais (SHECTMAN, 2009). Busca, também, incentivar a construção do pensamento crítico, a partir da relação entre professor e aluno (VARELA *et al.*, 2014), intermediada por ferramentas culturais e signos. Este tipo de mediação tem por enfoque a atuação do professor, elo entre o aprendiz e o conhecimento, a partir do diálogo, das experiências compartilhadas e da resolução de problemas (MASETTO, 2000). Os professores empreendemos o processo de mediação, por meio de instrumentos que tornam possível a interlocução entre si, o conteúdo e o aluno, entre estas ferramentas estão as tecnologias da informação e da comunicação – TIC, consideradas agentes de mudança, por propiciarem acesso à informação de diferentes modos, ampliando capacidades cognitivas (LÉVY, 1999).

A partir do compartilhamento de tecnologias entre os indivíduos, torna-se possível, também, o aumento do potencial de inteligência coletiva, uma vez que todos os envolvidos são postos a pensar e a desenvolver algo. Para Vygotsky (1981, p. 137),

> A introdução de uma nova ferramenta cultural num processo ativo, inevitavelmente o transforma. Nessa visão, recursos mediadores como a linguagem e as ferramentas técnicas não facilitam simplesmente as formas de ação que irão ocorrer, mas alteram completamente a estrutura dos processos mentais (VYGOTSKY, 1981, p. 137).

Dessa forma, se torna possível variar as práticas pedagógicas, uma vez que o professor se vê diante de um novo contexto de conhecimento. Segundo Behrens (2000), reconhecer a era digital como uma nova maneira de categorização do conhecimento não abre precedentes para o descarte do caminho trilhado, até então, pelo modelo tradicional, nem para a mitificação do uso exacerbado de computadores e afins no processo de ensino. Por outro lado, implica adequar o uso de ferramentas eletrônicas e/ou tecnológicas para construção dos processos metodológicos mais criativos, demandando do professor a possibilidade de ir além, ainda que permaneça dentro da sala de aula. Recursos tecnológicos assinalam meios de se instigarem novas metodologias, favorecendo a aprendizagem com interesse, criatividade e autonomia.

Pesquisas como as de Garcia (2016), Costa *et al.* (2015), Gehlen e Delizoicov (2012), Perrone (2018) e Oliveira (2016) atestam que o uso de recursos tecnológicos agrega resultados positivos no que tange ao alcance de novos conhecimentos. Em todas as pesquisas, ocorreu a mediação por parte de um recurso tecnológico. As duas últimas, especificamente, tratam do ensino de Física e Matemática, a partir da utilização da realidade aumentada.

## 3 FÍSICA NEWTONIANA: BREVE ABORDAGEM

As Leis de Newton constituem a base do que entendemos sobre o movimento e suas causas e suas limitações. No século XX, através da descoberta da Física quântica e da relatividade geral e específica, foram reveladas as limitações que existiam na lei do movimento de

Newton. A mecânica clássica foca no movimento do objeto particular que, ao interagir com os objetos que estão na sua vizinhança, tem sua velocidade acelerada (RAMOS, 1990).

A primeira Lei de Newton diz que não precisamos necessariamente de uma causa para um objeto se movimentar, ou seja, que é uma força que faz o objeto variar a sua velocidade. Assim, um objeto permanecerá em repouso ou em movimento uniforme, em linha reta, a menos que tenha seu estado alterado pela ação de uma força externa. Esta lei é também chamada de Lei da Inércia ou Princípio da Inércia. Em outras palavras, se um objeto não estiver em movimento, a propensão é de continuar em repouso e se está se movimentando vai continuar no mesmo movimento com a mesma velocidade se a força resultante for nula. O importante, na primeira lei, é conhecer a força resultante.

Podemos resumir a segunda Lei de Newton da seguinte forma, a força resultante que atua sobre um corpo é proporcional ao produto da massa pela aceleração por ele adquirida. As grandezas força gravitacional, força peso, força normal compõem a segunda lei.

De acordo com a terceira Lei de Newton, não é possível uma situação em que apenas uma única força atue isoladamente. A terceira Lei de Newton se resume a, quando um corpo 1 exerce uma força sobre o corpo 2, o corpo 2 exerce uma força sobre o corpo 1, essas duas forças são sempre iguais, em intensidade, e opostas, em sentido. Podemos indicar sinais diferentes para representação de sentidos opostos. Podemos rotular as forças como forças de ação e reação por causa da interação mútua dos corpos. Qualquer das duas forças poderá ser chamada de ação ou de reação, então esses rótulos devem ser arbitrários. A regra é que toda vez que tivermos uma ação, teremos uma reação igual, mas com sentido oposto.

## 4 TECNOLOGIA APLICADA AO ENSINO

As Tecnologias de Informação e Comunicação (TICs) são produtos que facilitam os métodos de ensino, atuando como ferramentas auxiliares no aprendizado de Física. Trazer aos alunos de Física a teoria, por meio de tecnologia, como o uso de Realidade Aumentada (RA) e Gifs, pode superar a falta de interesse, transformando-a em motivo para aprender.

Gif é a sigla para o termo inglês Graphics Interchange Format que significa formatos para intercambio gráfico, ou seja, é um formato de imagem que serve tanto para animações, quanto para imagens fixas. Quase sempre são atrelados ao humor e estão fortemente presentes em nosso cotidiano seja no *twiter*, no *facebook*, no e-mail ou em páginas da *web*. São utilidades na internet para comunicação rápida, são universais, por isso, quando escolhidos da maneira correta, transmitem mensagens a qualquer pessoa, em qualquer lugar. Eles geralmente são formulados para durar no máximo 8 segundos para captar a atenção do observador.

A busca por novas tecnologias que possam atrair nossos alunos também está expressa nos Parâmetros Curriculares Nacionais (PCN), que defendem uma alfabetização científica, a qual pode ocorrer por meio de:

> [...] novas e diferentes formas de expressão do saber da Física, desde a escrita, com elaboração de textos ou jornais, ao uso de esquemas, fotos, recortes ou vídeos, até a linguagem corporal e artística. Também deve ser estimulado o uso adequado dos meios tecnológicos, como máquinas de celular, ou das diversas ferramentas propicias pelos microcomputadores (BRASIL, 2017).

## 4.1 O QUE É REALIDADE AUMENTADA?

A realidade aumentada configura-se como a interação entre ambientes virtuais e o mundo físico, estabelecendo a interação entre elementos ou informações virtuais e a visualização do mundo real por meio de um aplicativo e uma câmera que proporciona uma experiência interativa com o mundo real, na qual os objetos do mundo real são "acentuados" por informações criadas por captadores virtuais. A realidade aumentada agrega o virtual ao ambiente natural (ROLIM *et al*. 2011).

A RA, através de Gifs no ensino das leis de Newton, pode auxiliar o professor na construção de conhecimento, ao ser apresentada por meio de símbolos, podendo atrair a atenção do aluno. Essa forma de apresentação de objetos e fenômenos traz a tecnologia para a realidade, possibilitando-nos viajar ao espaço em nossas próprias salas de aula e observar os planetas girando com a atração da gravidade em seu redor. Essa tecnologia permite-nos visualizar pequenos vídeos (Gifs) demonstrando fenômenos de nosso cotidiano, sendo possível adaptá-los para nossos livros didáticos com a proposta de utilizar a RA para ajudar o professor na inclusão da tecnologia e facilitar a comunicação (AZUMA, 1999). Nosso objetivo é implementar TICs, através de Gifs e RA, em uma revista em quadrinhos, onde os personagens serão baseados em *animes*, devido à grande aceitação dos jovens a esse modelo de revista. Propomos organizar uma revista em que os personagens dialogam, com linguagem simples, sobre as leis de Newton. A cada exemplo, teremos figuras em Gifs ou 3D, as quais são engatilhadas por um aplicativo de celular para leitura dos símbolos criados como transmissores da linguagem para que alunos observem, e sejam capazes de interpretar os fenômenos físicos das leis de Newton.

Esse aplicativo guardará dados de leitura dos símbolos, que serão programados de acordo com o assunto abordado e visualizado pelo programa. Através da tela do celular, o aluno poderá observar os exemplos em movimento, projetados na folha da revista figuras em Gifs ou em 3D, direcionados ao contexto estudado. Nossa intenção é que esse aplicativo seja colocado à disposição para ser baixado no Google Play e possa ser adquirido por qualquer pessoa em qualquer lugar. Ele pode ser acionado através de marcações existentes na revista. A inovação poderá ser utilizada por todo país sem utilização da internet, já que o aplicativo guardará as informações (Gifs e 3D), utilizados na revista. O aplicativo também pode ser baixado no blog do autor, *https://physics7.school.blog,* onde já está disponível.

## 5 APRENDENDO FÍSICA COM QUADRINHOS E TICs

Podemos considerar a revista em quadrinhos como uma atividade lúdica (RAMOS, 1990; HUIZINGA, 2001) com humor e regras para leitura. O acompanhamento do conteúdo programático articulado à forma lúdica de compreensão pode trazer um equilíbrio entre arte e ludicidade. Buscar através da catarse a libertação dos alunos das aulas tradicionais, retirando-lhes a tensão, pode fazer com que o ocorra aprendizagem, de forma criativa, por parte do discente (RAMOS, 1990). A atividade lúdica está inteiramente relacionada ao desafio. Desafiar o aluno a interpretar e mesmo reproduzir os vídeos implica trazer situações e obstáculos comuns do seu cotidiano para resolvê-los cientificamente.

## 6 PERCURSO METODOLÓGICO

Trata-se de uma pesquisa qualitativa, cujos dados foram colhidos no transcurso do desenvolvimento de uma sequência didática, na qual foram trabalhados conteúdos de mecânica clássica em uma turma de Ensino Médio, na cidade de Guarapari-ES.

6.1 OS OBJETOS VIRTUAIS NA APLICAÇÃO DA SEQUÊNCIA DIDÁTICA

O processo de criação e customização dos códigos inseridos na revista em quadrinhos traz uma forma diferenciada para pesquisa, pois assume um material de poucas referências iguais existentes no mercado. O desafio era confeccionar um material com linguagem jovem, resguardando conteúdo original para o ensino das Leis de Newton, usando, como atrativo, os códigos que ativam pequenos Gifs (série de imagens reproduzidos em um *loop* de tempo x, de uma determinada situação) e figuras de realidade aumentada com modelagem 3D.

Inicialmente, o protótipo da revista foi criado em PowerPoint e tem a sequência da história contada por figuras representativas retiradas de páginas na internet. Após esta parte da formulação, pesquisamos, junto com o responsável pela programação do aplicativo (Pap.), os Gifs que poderiam ser utilizados em cada situação e onde seriam inseridos na revista. Foi descrito ao desenvolvedor do sistema como deveria ser direcionado o aplicativo que leria os códigos e ativaria os signos existentes. O desenvolvedor, através da plataforma Unity, em conjunto com *plug-ins* (Vuforia) e algumas opções de configuração para realidade aumentada, adicionou às cenas uma série de planos e, nos mesmos, através de um código, utilizando a linguagem de programação *Java script* e a orientação a objetos, inserimos os Gifs. Os Targets serviram de gatilho para ativação dos códigos que exibem os Gifs na tela do dispositivo. A programação demorou quatro dias de trabalho e foi realizada por um estudante do curso de Sistemas de Informação da Universidade Federal do Espírito Santo.

6.2 OBJETOS VIRTUAIS DESENVOLVIDOS E ACIONADOS NA REVISTA

O aplicativo Ar Phisycs - Leis de Newton foi desenvolvido exclusivamente para essa pesquisa e está em processo de publicação para que fique disponível na Google Play Store, para ser baixado gratuitamente. O aplicativo também pode ser baixado no blog do autor, https://physics7.school.blog, onde já está disponível. Na primeira Lei de Newton, foi colocado um Gif onde uma pessoa é imobilizada em uma maca, em uma caminhonete aberta, sem nenhum tipo de objeto prendendo a maca. Quando ocorre o movimento, a tendência do corpo é permanecer parado ou em MRU, por isso ele cai do veículo.

Na segunda lei, são apresentados dois Gifs e um deles apresenta um pequeno cão de raça pug, que empurra um carrinho com um cãozinho de pelúcia dentro. O segundo Gif apresenta um senhor de idade avançada que estaciona o carro apenas o empurrando. Na terceira Lei de Newton, temos três Gifs, o primeiro descreve a decolagem de um foguete saindo do solo terrestre, o segundo dispõe de uma bala de revólver que se estilhaça na parede e o terceiro contém o encontro de ação e reação do rosto de um homem com um balão cheio de água com tinta. A revista ainda demonstra as grandezas: Força - com o Capitão América segurando um helicóptero, Força resultante, força de atrito, força peso, Força peso. Temos a representação das forças de interação à distância representada por uma figura 3D demonstrando a interação Lua x Terra e a interação por contato apresentado por um carro de Fórmula 1 atritando o solo. A força normal foi dividida em três figuras montadas juntas, com posições diferentes, visando a apresentar que a normal aparece no momento de encontro dos objetos. Consta, ainda, um Gif demonstrando a queda de uma mola e um Gif com movimento de um pêndulo.

6.3 CONTEXTO DE APLICAÇÃO DA SEQUÊNCIA DIDÁTICA

A aplicação da sequência didática ocorreu na escola de Ensino Médio Dr. Silva Mello na cidade de Guarapari, no estado do Espirito Santo, em uma turma de primeiro ano do Ensino Médio, escolhida aleatoriamente, no transcurso dos meses de outubro e novembro de 2018. A aplicação transcorreu no final do ano, devido à grande dificuldade da turma de primeiro ano com conteúdos básicos, sendo necessária uma ação de reforço matemático e aulas de leitura e interpretação, no primeiro trimestre do ano, atrasando o conteúdo de Física.

**7 RESULTADOS: APLICAÇÃO DA SEQUÊNCIA DIDÁTICA**

7.1 QUESTIONÁRIO DE CONHECIMENTOS PRÉVIOS

Antes de iniciar a aplicação da sequência didática (GIORDAN, 2005), foi apresentado aos alunos um questionário inicial para verificação do conhecimento prévio sobre as Leis de Newton, cujas respostas demonstraram, para a mesma pergunta, percepções diferentes tanto do conceito interpretado de seu cotidiano, quanto do ponto de vista científico. O questionário foi aplicado para uma turma de 1º ano de Ensino Médio com 35 alunos. Logo na primeira pergunta: "como você percebe se o corpo está parado ou em movimento?" 78% dos alunos responderam de forma errada, 10% de forma incompleta, apenas 9% de forma certa e 1% não soube responder. Nas demais perguntas, os alunos seguiram o mesmo padrão, no total foram 4% de acertos, 12% de respostas incompletas, 55% respostas erradas e 29% dos alunos deixaram em branco ou não souberam responder. Com base nesses dados, uma média de 67% não demonstrou conhecimento prévio para responder às perguntas do questionário inicial.

Julgamos necessário, antes da sequência didática, ministrar uma aula introdutória sobre conhecimentos prévios e uma aula mediada com recurso tecnológico através do *software* que trabalha com simulações com Tecnologia Educacional em Física (PHET), abordando conceitos básicos sobre massa e peso, porque houve muitas respostas erradas com relação à segunda Lei de Newton. Tópicos da Mecânica Clássica não foram tratados nessa intervenção. Revisamos os conteúdos Física básica, História da Física, notação científica, transformações, grandezas e medidas, representações vetoriais, operações com vetores, estudo do movimento, M.R.U. e M.R.U.V. Antes de iniciar a sequência didática, utilizamos o simulador PHET para os alunos experimentarem situações aplicadas aos conhecimentos adquiridos na aula anterior.

7.2 APLICAÇÃO MEDIADA DA PRIMEIRA LEI DE NEWTON

Os alunos foram separados em grupos de cinco e foi-lhes distribuída a revista, para leitura e representação de cada personagem. Foram distribuídos os quatro personagens para quatro alunos. Cada um foi lendo a sequência de frases representando seus personagens. Na introdução da revista, temos a história de um grupo de alunos que não gostam de Física e precisam fazer uma prova sobre as três leis de Newton. Os personagens se parecem com estudantes de Ensino Médio perdidos na matéria e acreditam que será muito difícil entender os conceitos. O texto é curto, de linguagem simples, voltada para o adolescente do século XXI.

Figura 1 – Página 1 da revista: a linguagem utilizada busca simular uma rede social

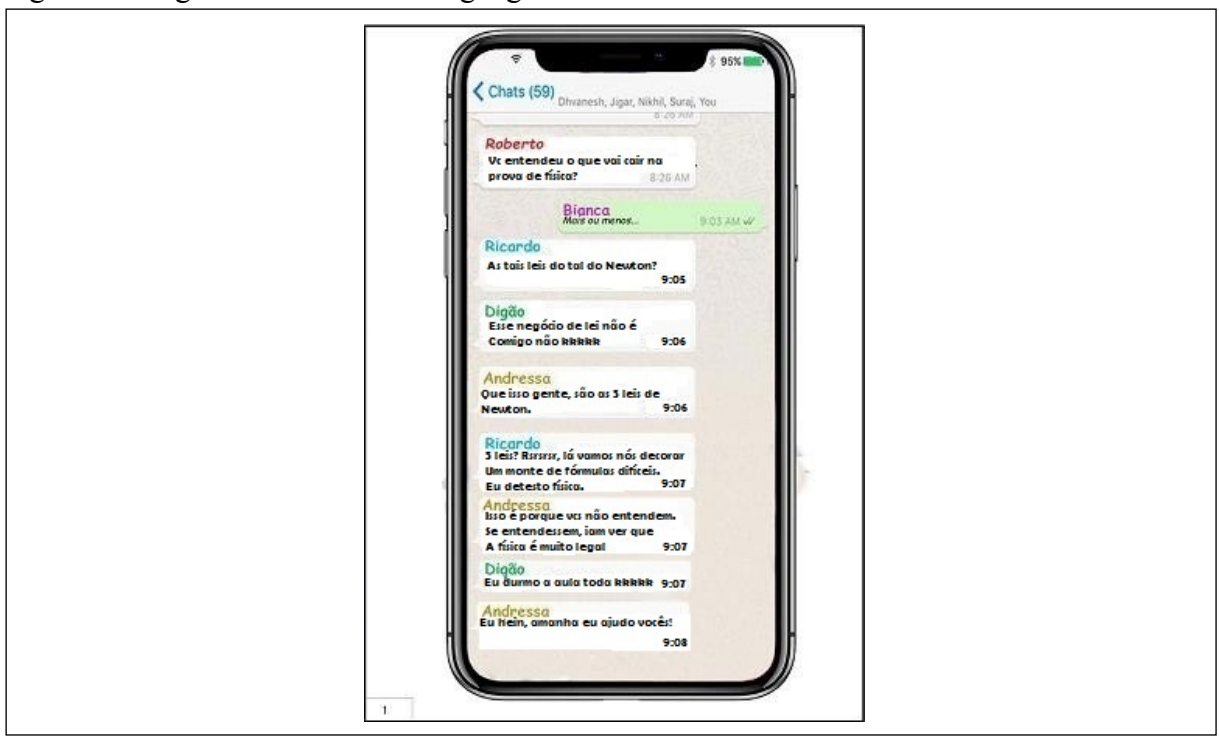

Fonte: Elaborado pela autora (2018)

A Figura 1 apresenta uma discussão em um grupo de aplicativo onde os alunos elegem suas dificuldades com relação às leis de Newton. Observando o teor das mensagens, uma colega se propõe a ajudar e agendam um encontro para estudar. Nas páginas subsequentes (Figura 2), ela aborda um pouco sobre a história de Isaac Newton e inicia a primeira Lei de Newton.

Figura 2 – Diálogo sobre a 1ª Lei de Newton na página 4 da revista

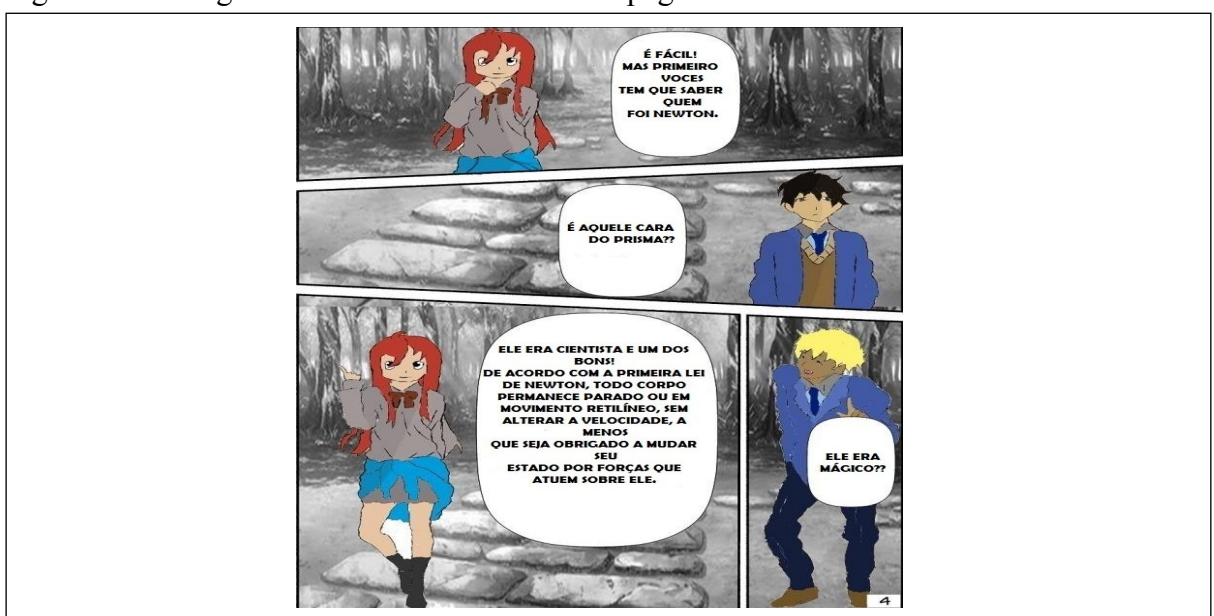

Fonte: Elaborado pela autora (2018)

A personagem Andressa expõe que estamos constantemente em contato com fenômenos físicos, apesar de não os percebermos. A partir da introdução, os alunos demonstraram interação com a revista. Ao final da leitura, instruímos os alunos que haveria um Target descrevendo a experiência, o qual deveria ser acionado no aplicativo AR Phisycs no telefone da professora, e analisar o Gif (Figura 3), um fenômeno real, para responder a uma pergunta diferente para cada grupo.

O aplicativo foi passado para os grupos, um de cada vez, os quais não deveriam se comunicar entre si. Os grupos foram instruídos a observar o Gif e elaborar uma pequena resenha para ser avaliada. Os grupos não podiam pedir, ou trocar, informações com os outros grupos. Além da resenha sobre o Gif, foi entregue uma pergunta para cada grupo. Havia sete perguntas que foram sorteadas e respondidas, cada uma, por um grupo diferente. Algumas perguntas pediam que o grupo respondesse e explicasse para toda a sala. Dos sete grupos, 6 chegaram a respostas satisfatórias e 1 respondeu de forma inconclusiva. As respostas indicaram ganho de conhecimento quando comparadas com as do questionário inicial.

Figura 3 – Aplicativo AR- Physics: p. 6, 1º Gif, primeira Lei de Newton, homem na maca

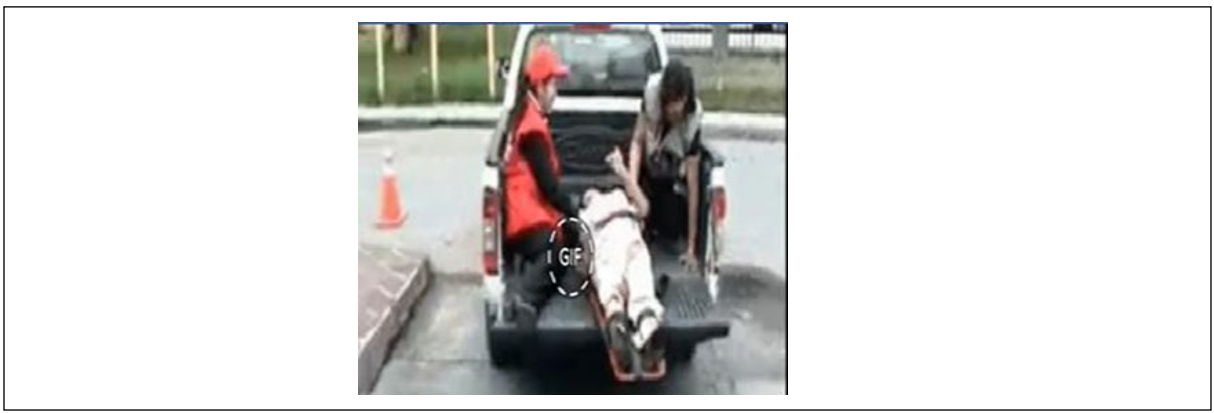

Fonte: WEBMODEVIDEOS (2012)

Após a análise, verificamos que todos os grupos descreveram o fenômeno de maneira próxima à científica. Não houve nenhum grupo que interpretou o Gif de maneira incoerente de acordo com o assunto abordado. Analisando a resposta do grupo 5 (Figura 4), observamos que as respostas começam a ganhar uma linguagem científica.

Figura 4 – Resposta do grupo 5

Transcrição: "Sim tendemos a permanecer em estado natural a não ser que haja outra força sobre nós e quando o trem entra em movimento nosso corpo tende a ser jogado para trás...1 Lei de Newton. Lei da inercia".

Fonte: Elaborada pela autora (2018)

A resposta do grupo 5 apresenta uma coerência perfeita, o desenvolvimento do texto descreve o fenômeno de forma objetiva e satisfatória. Apesar de ter iniciado a resposta com uma afirmativa que uma pessoa, no interior do trem é arremessada para frente, quando o trem anda, demonstrando força na inércia, logo após a afirmativa, descreve corretamente a inércia e os demais conceitos ligados a ela. Esse grupo descreveu o conceito científico subjacente à resposta, concernente à pergunta, para toda a turma, respondendo às dúvidas quando questionados. A professora, após o relato do grupo, fez os esclarecimentos necessários referente ao equivoco descrito pelo grupo 5 na resposta (Figura 5), para toda turma.

Figura 5 – Grupo 1 reunido para a dinâmica, em trabalho de equipe

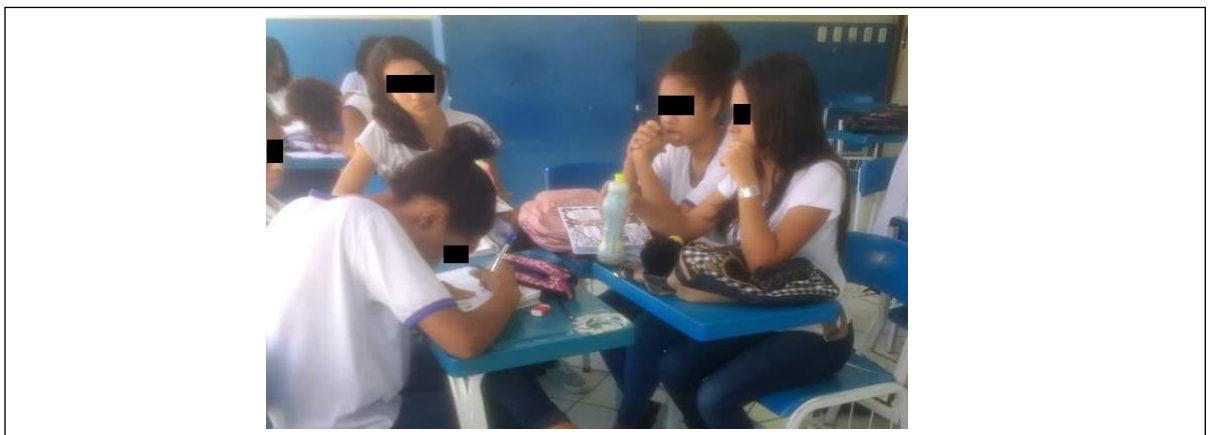

Fonte: Elaborada pela autora (2018)

## 7.3 SEGUNDA LEI DE NEWTON

Para a introdução da segunda Lei de Newton, seguimos com a leitura da revista. Os alunos mantiveram o mesmo processo, acompanhando, passo a passo, os quadrinhos em voz alta. Ao final da leitura, a estratégia de ensino se repetiu com a formação dos grupos e a entrega das perguntas que foram sorteadas, uma para cada grupo. Ao final, a professora novamente acionou o aplicativo AR Physics em seu celular, passado de grupo em grupo.

De acordo com o texto e a interpretação dos Gifs do cãozinho empurrando o carrinho (Figura 6) e o homem estacionando o carro, os alunos deveriam responder às perguntas e interpretar o fenômeno que cada um dos Gifs demonstrava, explicando-os de acordo com sua compreensão. Nessa dinâmica, foram utilizadas duas aulas geminadas de 55 minutos cada.

Na Figura 6, um cãozinho empurra um carrinho de bebê, exercendo uma força no carrinho, mantendo-o em movimento retilíneo e uniforme. A imagem foi escolhida para conduzir o olhar do aluno a perceber que a força de atrito contrabalanceia a força exercida pelo cãozinho, demonstrando a segunda Lei de Newton. As respostas dos alunos às perguntas demonstraram ganho de conhecimento e avanços nos fundamentos científicos. Os grupos conseguiram interpretar a pergunta de uma forma mais concisa, permitindo-nos detectar a utilização do conhecimento na resolução do problema, na relação entre força, massa e aceleração. Um grupo, porém, não conseguiu associar a direção da aceleração com a gravidade.

Figura 6 – Acionamento do AR-Physics: cão exercendo força sobre uma massa

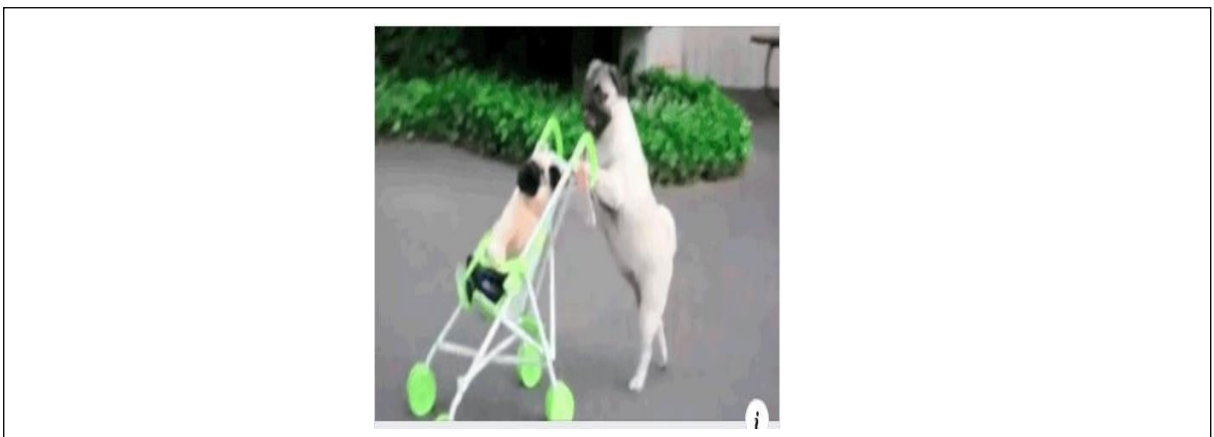

Fonte: WEBMODEVIDEOS (2012)

7.4 TERCEIRA LEI DE NEWTON

Na sequência, foi apresentada a terceira Lei de Newton a partir da revista e do acionamento de três Gifs. Por questão de escolha, neste artigo, apresentamos somente um gift. Os grupos voltaram a se reunir com os mesmos componentes com que trabalharam nas aulas anteriores. As perguntas sobre a terceira Lei de Newton foram cortadas em tiras e cada grupo elegeu uma para responder. A professora instruiu que eles deveriam discutir essas perguntas entre o grupo para formularem as respostas de acordo com a leitura das páginas indicadas. Apresentamos a terceira Lei de Newton com um exemplo sobre lançamento de foguetes e discutimos o Gif, onde constatamos o lançamento de um foguete (Figura 7) usando o princípio da ação e reação. Nesse momento, os alunos deveriam observar o lançamento e ler a explicação na revista.

Figura 7 – Acionamento do AR-Physics, 4º Gif, 2ª lei de Newton, foguete decolando

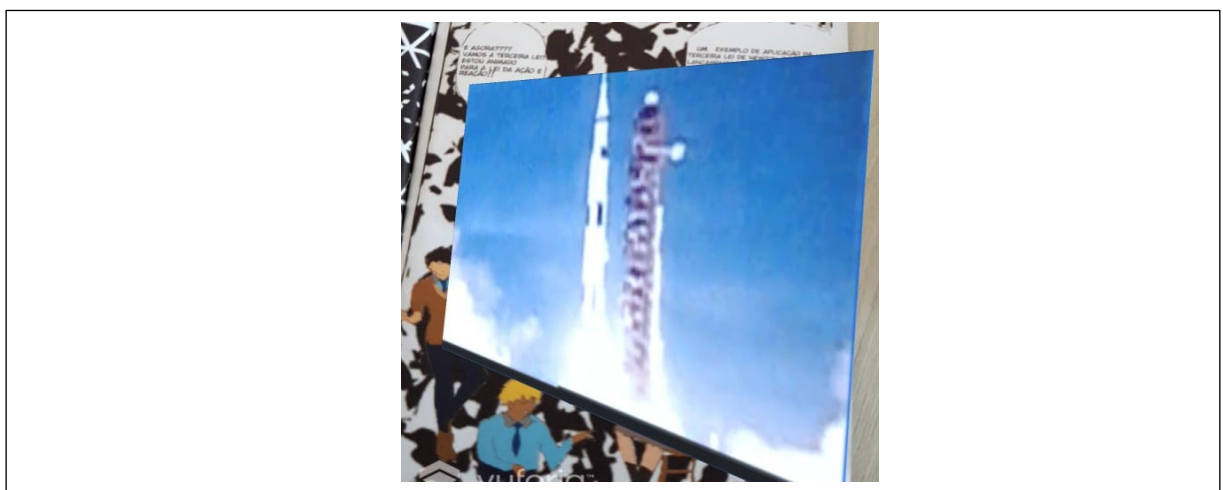

Fonte: HECRIPTUS (2018)

Após a apresentação do Gif (Figura 7), a professora solicitou a produção de um texto para relacionar o conteúdo da terceira Lei de Newton aos vídeos. Os textos produzidos pelos alunos demonstraram avanços no raciocínio, quando comparado ao questionário de conhecimentos prévios. De um modo geral, revelaram respostas bem elaboradas, indicando domínio de conteúdo, já que, nas perguntas da aplicação da primeira Lei de Newton, os grupos tinham dificuldades para interpretar e para organizar as respostas que deveriam ser formuladas. Os alunos utilizaram dados da revista para interpretar o Gif (Figura 7), para redigir o texto pedido. As respostas dos grupos até aqui apresentadas trazem conteúdos coerentes, explicitando o movimento dos corpos distintos. As respostas, com o transcurso da sequência didática, tornaram-se mais científicas, o que demonstra uma diferença do entendimento das forças de ação e reação apresentado no primeiro questionário de conhecimentos prévios. Se observarmos as respostas apresentadas, verificaremos que o conhecimento científico, embora incompleto, aparece sempre nas respostas, sinalizando avanços na alfabetização científica.

A aplicação da dinâmica da terceira Lei de Newton também durou 2 aulas de 55 minutos cada (8ª e 9ª aplicações). Ao final da segunda aula (9ª aplicação), foi escolhido um grupo para falar sobre a pergunta sorteada e sobre a interpretação do Gif à escolha do grupo.

Após as leis de Newton, iniciamos o estudo do movimento e suas causas e abordamos sobre o movimento dos corpos, buscando as causas e relacionando-as com suas consequências, para aplicações no cotidiano do aluno. Nessa etapa, os grupos analisaram, e discutiram, em partes, respondendo a perguntas e fazendo relatórios sobre os novos Gifs visualizados. Após o estudo das três leis de Newton, então, abordamos várias grandezas, entre as quais, força resultante, força peso, momento em que enfatizamos o estudo do conceito de campo gravitacional e sua atuação na dinâmica dos corpos, no processo de queda livre. Por uma questão de tempo e de escolhas, optamos por não apresentar todo o processo, evitando que o artigo se alongasse e fugisse do escopo da revista.

## 8 ANÁLISE DOS RESULTADOS: APLICAÇÃO DE QUESTIONÁRIO PARA VERIFICAÇÃO DE APRENDIZADO:

Nos resultados finais, observamos avanços na utilização de conceitos científicos. Constatamos que, nos acertos do questionário final, de 386 perguntas respondidas, tivemos 217 respostas corretas, 38 respostas incompletas, 120 respostas erradas e 11 respostas em branco. No primeiro questionário, os alunos deixaram muitas questões em branco, o que diminuiu no questionário final. Em geral, percebemos ganho de conceitos científicos, o que ficou evidenciado na linguagem apresentada nos textos redigidos. A época escolhida para aplicação ocorreu quando os alunos já estavam cansados e, praticamente, se preparando para as férias de fim de ano, isso enriquece o resultado, pois, geralmente, não se consegue atrair muito a atenção do aluno nesse período, mas a sequência didática fez com que eles adquirissem conhecimento, fossem participativos e colaborativos. A sequência didática aplicada tinha o objetivo de construir conhecimento de forma lúdica, o que apresentou um bom resultado, se observados os índices do questionário final. O índice apresentado no Gráfico 1 da avaliação final apresenta ganho de conhecimento nas respostas consideradas corretas, pois, no questionário de conhecimentos prévios, a maioria das respostas (86%) não revelava conhecimentos científicos sobre mecânica clássica por parte dos alunos.

Gráfico 1 – Verificação de conhecimentos questionário final

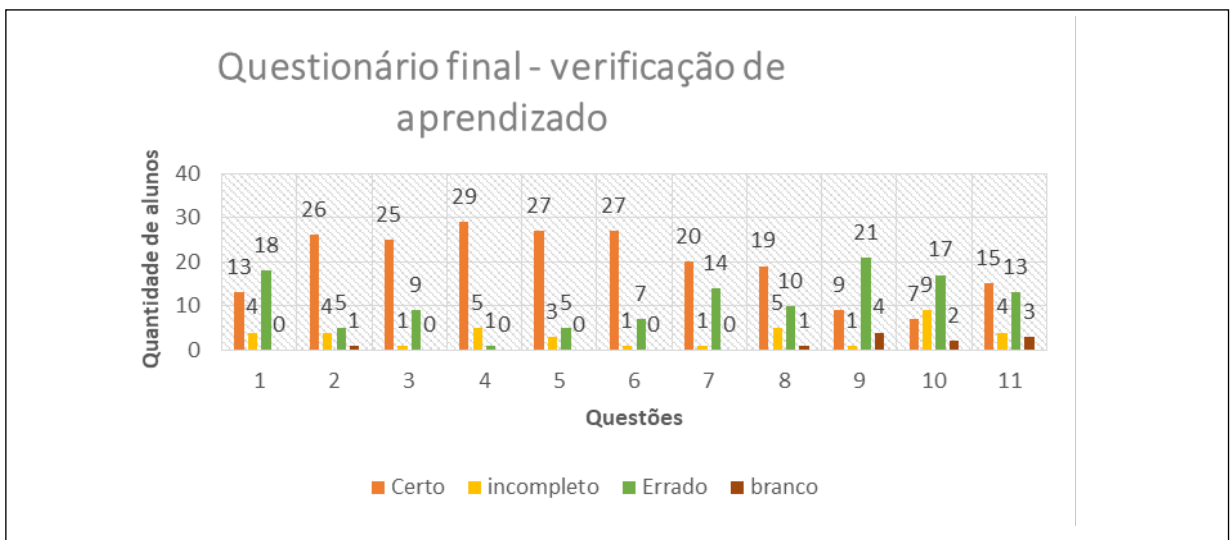

Fonte: Elaborado pela autora (2018)

Observando o Gráfico 1, temos um aumento considerável de acertos e de respostas erradas e os casos de respostas em branco praticamente desapareceram no gráfico de avaliação final, com apenas 5 eventos. O índice de respostas incompletas também diminuiu. O Gráfico 2 apresenta a comparação entre as respostas apresentadas no questionário inicial e as apresentadas no questionário final.

Gráfico 2 – Comparação de resultados questionário inicial x questionário final

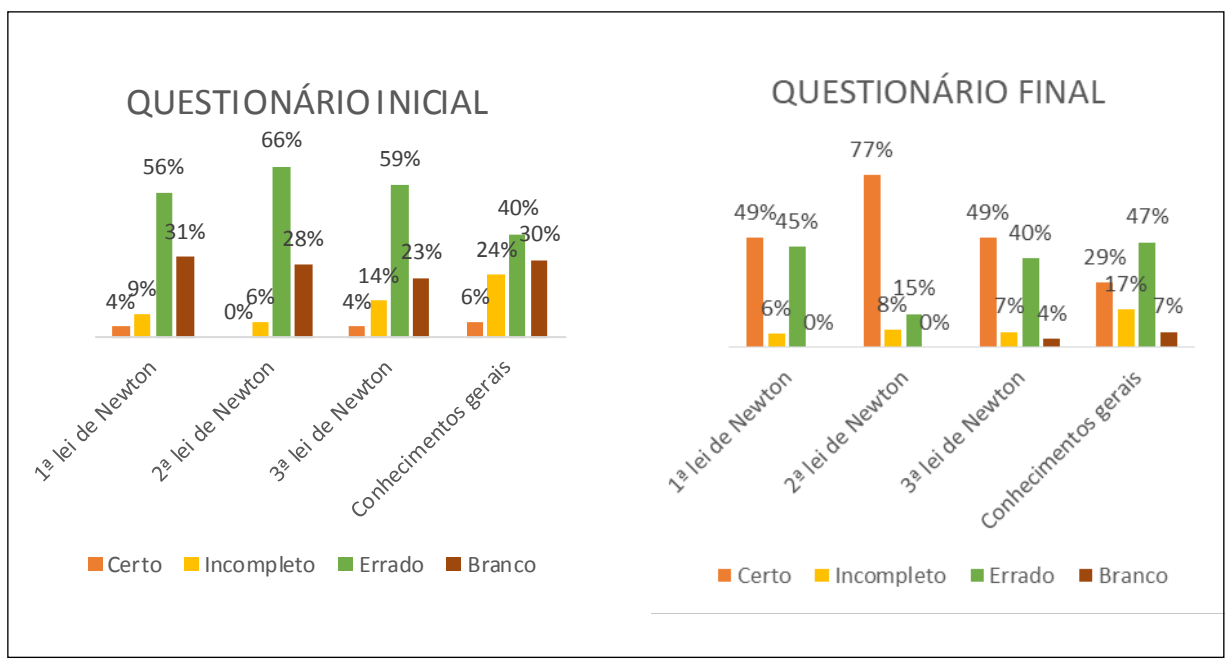

Fonte: Elaborado pela autora (2018)

Podemos observar que no primeiro questionário, no resultado da coluna referente à primeira Lei de Newton, tivemos 56% de perguntas respondidas de forma errada e 31% em branco, ou seja, se contarmos a soma das duas colunas, teríamos 87% das questões respondidas sem embasamento científico. Tivemos 13% perguntas apenas, respondidas de forma a articular a resposta ao conceito científico. Concluímos que, no início da aplicação, a maioria dos alunos não possuía conhecimento científico sobre a primeira Lei de Newton. Se compararmos com a coluna das respostas sobre a primeira Lei de Newton, no questionário final, temos 0% de respostas em branco e 45% erradas, o que apresenta um índice muito menor se comparado com as questões respondidas anteriormente. Analisando o índice de conhecimento científico no questionário final, temos, na coluna da primeira Lei de Newton, 49% de respostas corretas e 6% incompletas, somando o total de 55%, indicando ganhos no conhecimento. Isso revela um índice bem maior nas respostas científicas, revelando avanços na alfabetização científica sobre a primeira Lei de Newton, a qual aponta para a capacidade de o aluno interpretar fatos e fenômenos sociais e naturais, à luz da ciência (SASSERON; CARVALHO, 2008).

Na segunda coluna temos as respostas referentes à segunda Lei de Newton. No questionário de conhecimentos prévios, esta é a barra que apresenta maior número de questões (66%) erradas. O índice de respostas em branco também é grande, temos 28% respostas em branco. Se compararmos com o questionário final, temos a maior barra da aplicação com 77% de questões, só que, no segundo questionário, essas respostas estão corretas. Além de as respostas corretas aumentarem, existem as respostas em branco, que também ficaram em 0%, as incompletas foram de 8% e as erradas diminuíram muito, passando a apenas 15%. Esse resultado nos permite concluir que os alunos auferiram ganho conceitual da segunda Lei de Newton, apresentando o melhor resultado entre os 4 módulos analisados.

Com relação à terceira Lei de Newton, o primeiro gráfico (Gráfico 1) apresenta um índice alto de respostas erradas e em branco, com um total de 59% de erradas e 23% em branco. Já no questionário final, temos 40% erradas e 4% em branco, os dois índices diminuíram no questionário final e os índices de acerto aumentaram de 4%, no primeiro questionário, para 49% no último questionário. O índice de respostas incompletas também diminuiu de 14%, no primeiro questionário, para 7%, no segundo questionário. Essa análise comparativa nos permite concluir que houve ganhos de conhecimento na terceira Lei de Newton.

Nas perguntas que apresentavam conhecimentos gerais ligados a conceitos aprendidos na sequência didática, houve melhora significativa. No gráfico inicial, verificamos 6% de respostas certas, já no gráfico final, temos 29% respostas certas. As respostas em branco também evidenciaram uma melhora de 30% no questionário inicial, para 7% no questionário final. Somente não tivemos melhoria de resultado no questionário final das respostas erradas. No questionário inicial, tivemos 40% e, no final, tivemos 47%. Dados assim nos ajudam a compreender que o aumento de respostas erradas, nesse último item, está relacionado com a migração de respostas em branco para respostas erradas, já que os alunos que não tinham conhecimento nem para tentar responder, no questionário inicial, poderiam ter respondido de forma errada no questionário final, já que o índice de respostas em branco reduz mais da metade, na comparação dos questionários.

Observando os gráficos e os avanços alcançados pelos alunos, inferimos que a aprendizagem do conteúdo foi gradual. Apesar de os resultados terem sido muito significativos, podemos apontar avanços na alfabetização científica no que diz respeito à utilização de termos e conceitos científicos. A sequência didática foi satisfatória, demandando alguns ajustes nas perguntas apresentadas e nos Gifs escolhidos.

Estabelecendo uma comparação dos resultados do questionário inicial para o questionário final, observamos algumas diferenças significativas. Podemos comparar os acertos do primeiro questionário compreendendo 4% do total de questões, no questionário final, temos 51% de acertos do total das questões apresentadas: o índice de respostas incompletas diminuiu de 13% no primeiro questionário para 10% no último questionário. Quanto às respostas erradas, temos 55% no primeiro questionário e 37% no último. Finalmente, nas respostas em branco, houve uma diferença expressiva de 28% no primeiro questionário para 3% no último.

Os melhores resultados apontam para as respostas certas, que aumentaram em 50% e, nas respostas em branco, as quais, no primeiro questionário, representaram 29% e, no último, 3%. As respostas erradas, apesar de diminuírem, são consideradas altas. As melhorias que devem ser feitas para futuras aplicações devem considerar esse índice para adaptação e aperfeiçoamento do material. Outra evidência que o Gráfico 2 indica é que os alunos que antes não tinham interesse em responder ao questionário, ou não souberam dar as respostas, passaram a participar ativamente, se observarmos os índices do segundo gráfico, mesmo que no índice de erros, mostrando, ainda que timidamente, interesse pela matéria.

## 8.1 AVALIAÇÃO DA PRÁTICA PEDAGÓGICA

Como avaliação da prática pedagógica, foram feitas 7 perguntas sobre a metodologia aplicada referente aos conceitos aprendidos na revista de realidade aumentada. Foram distribuídos 35 questionários aos alunos que participaram ativamente da aplicação da sequência didática. Apresentamos a seguir uma breve análise das respostas (PERRONE, 2018). Na primeira questão, ao perguntarmos se o conteúdo restou compreendido de maneira clara e objetiva, 85% responderam que sim e 15% responderam ter ficado com dúvida em algum dos conteúdos. Ao serem questionados se enfrentaram dificuldades com a metodologia, 94% avaliaram a metodologia fácil de entender e gostaram, mas 6% julgaram-na com nota mediana. A terceira pergunta buscava saber se gostariam de mais atividades como essas. Nesse caso, 100% dos alunos responderam que sim, mesma resposta que os alunos apresentaram para a quarta pergunta, que indagava se com a revista, eles conseguiram aprender mais. A quinta pergunta tratou de entender qual conceito e Gif foram melhor explicados e 42% responderam que gostaram do Gif que explicou a primeira Lei de Newton, 11% gostaram dos Gifs que explicaram a segunda lei, 28% avaliaram que os Gifs que representaram a terceira lei foram mais explicativos e os outros 19% dividiram-se entre as demais Gifs da revista.

A sexta pergunta buscava saber qual dos Gifs exigiu maior esforço de interpretação e, nesta questão, 6% dos alunos tiveram dificuldades de entendimento no Gif do cãozinho pug (segunda Lei de Newton), 6% no Gif que explode a bala na parede (terceira Lei de Newton), 14% no Gif do lançamento do foguete, 26% no Gif da queda da maçã (primeira Lei de Newton), 6% no Gif onde um senhor empurra o carro (segunda Lei de Newton), 6% tiveram dificuldade na interpretação do Gif do Capitão América, 6% no Gif final da demonstração da ação de uma mola para introdução da Lei de Hooke. Dez por centro não teve dificuldade nenhuma.

Na sétima pergunta, pedimos sugestões gerais para aprimoramento, 85% dos alunos disseram não haver sugestões, vez que aprenderam muito com a revista. Houve algumas sugestões tais como: melhorar os gatilhos ou *targets* de acionamento dos Gifs; fazer essa atividade fora da sala de aula; usar o aplicativo em todos os celulares da sala e usar a metodologia nas provas. Foi sugerido, também, que se expandisse a metodologia para outros conteúdos.

## 11 CONCLUSÃO

Nesse trabalho propusemos uma sequência didática mediada sobre as Leis de Newton utilizando uma revista que apresentava *targets* que direcionavam o acionamento de Gifs através do uso de um celular. Na análise do primeiro questionário de verificação de conhecimentos prévios do assunto abordado, constatamos que os alunos, ao ingressarem no Ensino Médio, não trazem praticamente nenhum conhecimento sobre elementos básicos para o aprendizado das leis de Newton. A revista ajudou na abordagem sobre as três leis de Newton, mostrando-se aliada da ação docente. Essa estratégia pode ser melhor trabalhada com uso constante, em sala de aula, adaptando-se ao contexto onde será feita a aplicação.

A sequência didática, aplicada com a ajuda da revista, teve uma boa aceitação entre os adolescentes, razão pela qual ajudou no aprendizado. As sugestões dos alunos podem melhorar de forma a ampliar a identificação do adolescente com o material, o qual pode ser testado em outros conteúdos. Desde a aplicação em 2018, já utilizamos o material em 7 turmas em 2019, sendo 2 turmas de EJA (Educação de Jovens e Adultos). Por perceber a grande aceitação do material, pretendemos utilizar em mais 7 turmas no ano de 2020 e ir procedendo às adaptações necessárias, de acordo com as observações em cada aplicação.

# REFERÊNCIAS